\documentclass[sigconf]{acmart}
\setcopyright{none}
\renewcommand\footnotetextcopyrightpermission[1]{} 
\usepackage{subfigure}
\usepackage{xspace}
\usepackage{cleveref}
\usepackage{hyperref}
\usepackage{xcolor}
\usepackage{bbding}
\usepackage{verbatim}
\usepackage{amsmath}
\usepackage{graphicx}
\usepackage{enumitem}

\AtBeginDocument{%
  \providecommand\BibTeX{{%
    \normalfont B\kern-0.5em{\scshape i\kern-0.25em b}\kern-0.8em\TeX}}}

\newcommand{\xw}[1]{\textcolor{black}{#1}}
\begin{document}
\pagestyle{plain}




\title{Exploiting Cross-Layer Vulnerabilities: Off-Path Attacks on the TCP/IP Protocol Suite}

\author{
  {Xuewei Feng}\textsuperscript{*},
  {Qi Li}\textsuperscript{*\ddag},
  {Kun Sun}\textsuperscript{\dag},
  {Ke Xu}\textsuperscript{*\ddag \Envelope},
  {and Jianping Wu}\textsuperscript{*\ddag},
  \\
  {\textsuperscript{*}Tsinghua University}, \textsuperscript{\ddag}Zhongguancun Laboratory, \textsuperscript{\dag}George Mason University\\
    \href{mailto:fengxw06@126.com}fengxw06@gmail.com, \{qli01@, xuke@\}tsinghua.edu.cn, ksun3@gmu.edu, jianping@cernet.edu.cn
}
\begin{abstract}

After more than 40 years of development, the fundamental TCP/IP protocol suite, serving as the backbone of the Internet, is widely recognized for having achieved an elevated level of robustness and security.
Distinctively, we take a new perspective to investigate the security implications of cross-layer interactions within the TCP/IP protocol suite caused by ICMP error messages. Through a comprehensive analysis of interactions among Wi-Fi, IP, ICMP, UDP, and TCP due to ICMP errors, we uncover several significant vulnerabilities, including information leakage, desynchronization, semantic gaps, and identity spoofing. 
These vulnerabilities can be exploited by off-path attackers to manipulate network traffic stealthily, affecting over 20\% of popular websites and more than 89\% of public Wi-Fi networks, \xw{thus posing risks to the Internet}. By responsibly disclosing these vulnerabilities to affected vendors and proposing effective countermeasures, we enhance the robustness of the TCP/IP protocol suite, \xw{receiving acknowledgments from well-known organizations} such as the Linux community, the OpenWrt community, the FreeBSD community, Wi-Fi Alliance, Qualcomm, HUAWEI, China Telecom, Alibaba, and H3C.
\end{abstract}




\maketitle

\section{Introduction}
\label{sec:intro}

\begin{figure}[h]
	\vspace{-3mm}
	\begin{center}
		\includegraphics[width=0.38\textwidth]{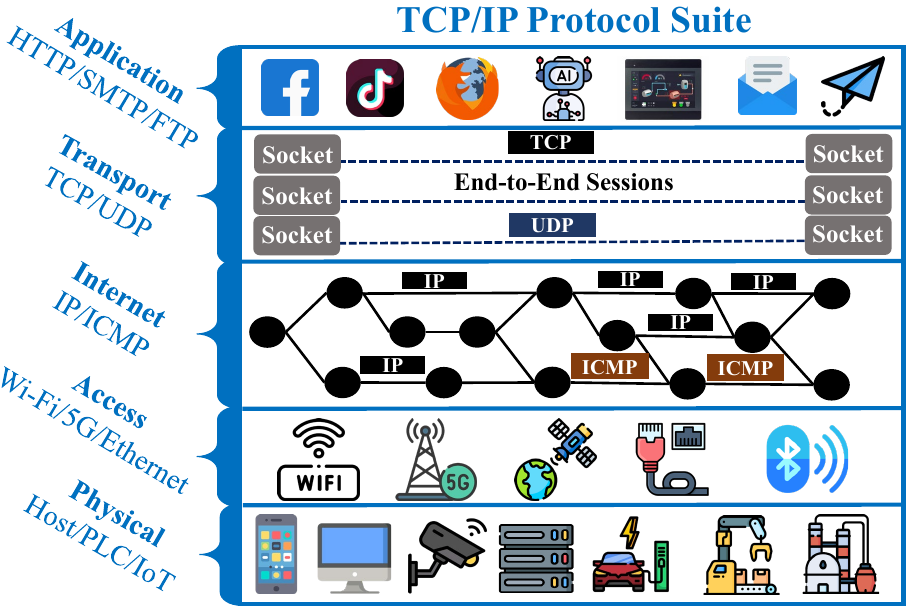}
		\vspace{-2mm}
		\caption{The TCP/IP protocol suite serves as the essential framework for data transmission on the Internet.}
		\label{pic:background}
	\end{center}
	\vspace{-3mm}
\end{figure}

The TCP/IP protocol suite is a set of communication protocols that underpin the Internet.
As shown in Figure~\ref{pic:background}, protocols at different layers of the suite (e.g., Wi-Fi, IP, TCP, and HTTP) form the essential framework for data transmission on the Internet.
Given the paramount significance of the TCP/IP protocol suite, it becomes a pivotal target for a myriad of attacks~\cite{ccsfeng,feng2022ndss,bellovin2004look,man2021dns,ccsman,gilad2011fragmentation,gilad2013fragmentation}. Exploiting and compromising the TCP/IP protocol suite can lead to extensive repercussions, posing a fundamental threat to Internet security and presenting significant incentives to attackers. To combat the diverse spectrum of network attacks, \xw{both industry and academia have dedicated substantial efforts~\cite{rfc7414,rfc6056,rfc5927,rfc5961,cao2016off,gilad2011fragmentation,qian2012off}}. 
\xw{However, in this paper, we demonstrate that vulnerabilities arising from cross-layer interactions among various protocols within the TCP/IP protocol suite, caused by forged ICMP (Internet Control Message Protocol) error messages, have received limited attention. These vulnerabilities can be exploited by off-path attackers, posing risks to the Internet.}


%

%

%

In the process of network data processing, protocols within the suite must interact and coordinate across layers. This cross-layer interaction ensures the smooth generation, transmission, and reception/storage of data. \xw{For example, when delivering an HTTP message, protocols such as DNS, TCP, IP, ARP, and Wi-Fi may need to be invoked to process and encapsulate the message.}
Although each protocol within the protocol stack may individually possess sufficient robustness, combining these protocols and engaging in cross-layer interaction through function calls can introduce security issues or anomalies. Specifically, the proper execution of one layer's specific functionality can be compromised by the normal execution of other layers. For instance, the loss of wireless frames in wireless networks commonly occurs due to inevitable communication noise interference; however, at the TCP layer, if TCP segments are not promptly acknowledged due to the loss of wireless frames, it can mistakenly trigger the detection of network congestion, \xw{leading to inefficient execution of the TCP congestion control algorithm~\cite{7164323}.}

\xw{In particular, ICMP, recognized as a fundamental component of the TCP/IP protocol suite, frequently drives cross-layer interactions that transcend traditional network layer boundaries to report network conditions or errors. By operating directly on top of IP, ICMP error messages embedded with various payloads can influence the behavior of higher layers such as TCP and UDP, and can even be exploited by off-path attackers to compromise higher-layer protocols.
In this paper, we undertake a comprehensive study to investigate the cross-layer interactions within the TCP/IP protocol suite caused by ICMP errors. Consequently, we uncover multiple vulnerabilities, including information leakage, desynchronization, semantic gap, and identity spoofing. These vulnerabilities can be triggered by forged ICMP errors issued from off-path attackers on the Internet, leading to exceptions during interactions among various protocols, such as Wi-Fi, IP, ICMP, UDP, and TCP. As a result, they pose significant risks to Internet security.}

\noindent \textbf{Information Leakage Leading to Off-Path TCP Hijacking.}
We delve into the interactions among ICMP, IP, and TCP and reveal that the IP Identification (IPID) field of the IP protocol, even with the most advanced IPID assignment policy currently available in Linux systems, can be manipulated by a forged ICMP error message.
This manipulation results in information leakage, exposing the confidential sequence number of the upper-layer TCP protocol. Consequently, off-path attackers on the Internet can craft acceptable TCP packets carrying the identified sequence number to poison or terminate the target TCP connection~\cite{ccsfeng,feng2021off}.
In this scenario, an off-path attacker on the Internet can impersonate an intermediate router by IP spoofing~\cite{ali2007ip}. Then the attacker issues a forged ICMP error message to a victim Linux server, thereby manipulating the server's IPID assignment policy for outgoing TCP packets.
This manipulation of the IPID value establishes an exploitable side channel for information leakage. By observing the altered IPID, the off-path attacker can deduce the random sequence and acknowledgment numbers of the upper-layer TCP connections between the server and victim clients.
\xw{Once these details are identified for a victim TCP connection, the off-path attacker can inject crafted out-of-band TCP packets into that connection, effectively achieving remote manipulation of the connection.}
Extensive measurements on the Internet reveal that over 20\% of popular websites are susceptible to the identified attack. The Linux community has acknowledged this vulnerability, assigning it the CVE identifier CVE-2020-36516. 

%

\noindent \textbf{Desynchronization Leading to TCP Traffic Poisoning.}
Another security vulnerability arising from cross-layer interactions within the TCP/IP protocol suite due to forged ICMP errors is the desynchronization issue. During interactions among IP, ICMP, and TCP, synchronization issues will arise in operations involving shared variables between these protocols~\cite{feng2022ndss}.
By crafting an ICMP error message, off-path attackers on the Internet can exploit this issue to force a target server's TCP packets to undergo undesired IP fragmentation. 
\xw{
This manipulation effectively circumvents the Path MTU Discovery (PMTUD) mechanism~\cite{rfc8201,rfc1191}, which is designed to prevent IP fragmentation of TCP packets.}
Subsequently, the attacker can pretend to be the server via IP spoofing and inject crafted IP fragments into the target TCP stream. This results in erroneous reassembly of the attacker's malicious fragments and the legitimate fragments from the server at the client's end, ultimately poisoning the messages received by the client.
Through extensive evaluations in the real world, we demonstrate that these attacks can be conducted to poison web cache, intercept HTTP redirection, and hijack BGP routing.

\noindent \textbf{Semantic Gap Leading to Routing Manipulation.}
We uncover a vulnerability stemming from the semantic gap in the validation of ICMP error messages, which can be exploited to manipulate the routing of a victim server, thus affecting all upper-layer traffic handled by IP~\cite{feng2022off-redirect}.
These flaws arise from insufficient consideration given to the generation of ICMP error messages caused by stateless protocols like UDP, ultimately resulting in a semantic gap. As a result, servers are unable to discern the legality of received ICMP errors carrying the payload of stateless protocol data\footnote{\xw{According to ICMP specifications~\cite{rfc792,rfc1122,rfc1812}, error messages should include at least the first 28 octets of the original packet to aid in identifying the affected process and verifying legitimacy. However, if the ICMP error message contains stateless protocol data payload, it becomes challenging for the receiver to verify legitimacy (details on ICMP errors in \S\ref{subsec:icmpbasics}).}}.
In this scenario, an off-path attacker on the Internet can forge an ICMP redirect message containing embedded UDP data, bypassing a target server's validation of message legality.
Consequently, the server mistakenly accepts the forged message, leading to manipulation of its IP routing (i.e., \xw{setting the IP address specified by the attacker in the ICMP redirect message as its next hop gateway}). \xw{This manipulation causes the server to direct its outbound traffic for the client towards its neighboring host (the specified next hop gateway) without the capability of traffic forwarding (i.e., a routing blackhole),} ultimately resulting in a Denial-of-Service (DoS) attack.
Through extensive measurements, we reveal that over 97,500 servers in 185 countries worldwide are vulnerable to these remote DoS attacks.

\noindent \textbf{Identity Deception Leading to Wi-Fi Hijacking.}
We trace the interactions among ICMP, Wi-Fi, and IP and identify an identity deception vulnerability caused by forged ICMP errors. This vulnerability allows a malicious client within a Wi-Fi network to impersonate the Access Point (AP) router and intercept wireless traffic from other clients~\cite{feng2022man}.
Specifically, the malicious client exploits IP spoofing to impersonate the AP router, sending fake AP-specific ICMP errors to deceive other clients into viewing the attacker as a better next-hop router.
As a result, the victim client sends decrypted wireless traffic directly to the attacker, bypassing the link-layer security mechanisms (e.g., WPA3) in Wi-Fi networks.
The root cause of the vulnerability stems from a design flaw in the Network Processing Units (NPUs) by manufacturers like Qualcomm and HiSilicon, which lack secure auditing of ICMP error-triggered cross-layer interactions.
Consequently, in our investigations, all 55 popular AP routers from 10 renowned vendors equipped with vulnerable NPU chips are unable to prevent malicious clients from impersonating AP routers and issuing forged ICMP errors.
Moreover, our measurements show that over 89\% of real-world Wi-Fi networks (e.g., Wi-Fi networks in coffee shops, hotels, campuses, or shopping malls) are vulnerable. The Wi-Fi Alliance, Qualcomm (which has assigned the CVE identifier CVE-2022-25667 to the vulnerability in its NPU), HUAWEI, H3C, and others have acknowledged our contributions to enhancing Wi-Fi security.

\section{Background and Threat Model}
\label{sec:IPID}

\subsection{Basics of ICMP Error Messages}\label{subsec:icmpbasics}
\xw{
ICMP error messages represent specific types of ICMP messages generated in response to network issues. They play a crucial role in identifying and diagnosing problems within a network~\cite{rfc792,rfc1122,rfc1812}.
These messages include Destination Unreachable (indicating network failures or host unavailability), Time Exceeded (resulting from packet Time-To-Live expiration), Parameter Problem (addressing IP header parameter issues), and Redirect Messages (optimizing routing). Additionally, Source Quench messages, historically significant for congestion control, are now deprecated.
Aiming to report network issues to the receiver, ICMP error messages inevitably induce cross-layer interactions within the TCP/IP protocol stack and prompt the receiver to adjust its behavior based on the received ICMP error messages.
According to the ICMP specifications~\cite{rfc792,rfc1122,rfc1812}, ICMP error messages should contain at least the first 28 octets of the original packet that triggered the error message (i.e., 20 octets of the IP header plus at least the first 8 octets).
When an ICMP error message is received, the receiver can utilize the embedded payload in the message to match it to the corresponding process. This enables the process to adapt and respond effectively.
For example, when an ICMP Destination Unreachable message with the code `Packet too big' is received, it facilitates cross-layer interactions by enabling the receiver's TCP to reduce its MSS (Maximum Segment Size), thereby avoiding IP fragmentation on the intermediate routes that issued the ICMP error message.
}

\xw{
Unfortunately, in practice, it is easy for attackers on the Internet to forge ICMP error messages to manipulate the receiver's behavior. Firstly, because ICMP error messages can be generated by any intermediate router along the network path, it is difficult for the receiver to authenticate the source of these messages. This is particularly challenging because attackers can use IP address spoofing techniques to forge the source IP address. Secondly, although ICMP specifications require that error messages include at least the first 28 octets of the original packet, enabling the receiver to match the message and perform a legitimacy check, attackers can easily forge a 28-octet payload to bypass this check.
}
\xw{
In the context of TCP communication, the first 28 octets of the original packet contain a random sequence number, which is hard for attackers to guess. However, in UDP or ICMP scenarios, since these protocols are stateless and lack randomized sequence numbers, attackers can easily forge a 28-octet payload to include in the falsified ICMP error message. This allows them to evade the receiver's legitimacy check, deceive the receiver into responding to the message, and lead to unintended protocol interactions, thereby posing security risks.
}

\subsection{Threat Model of Off-Path Attacks}
\begin{figure}[h]
	\vspace{-3mm}
	\begin{center}
		\includegraphics[width=0.35\textwidth]{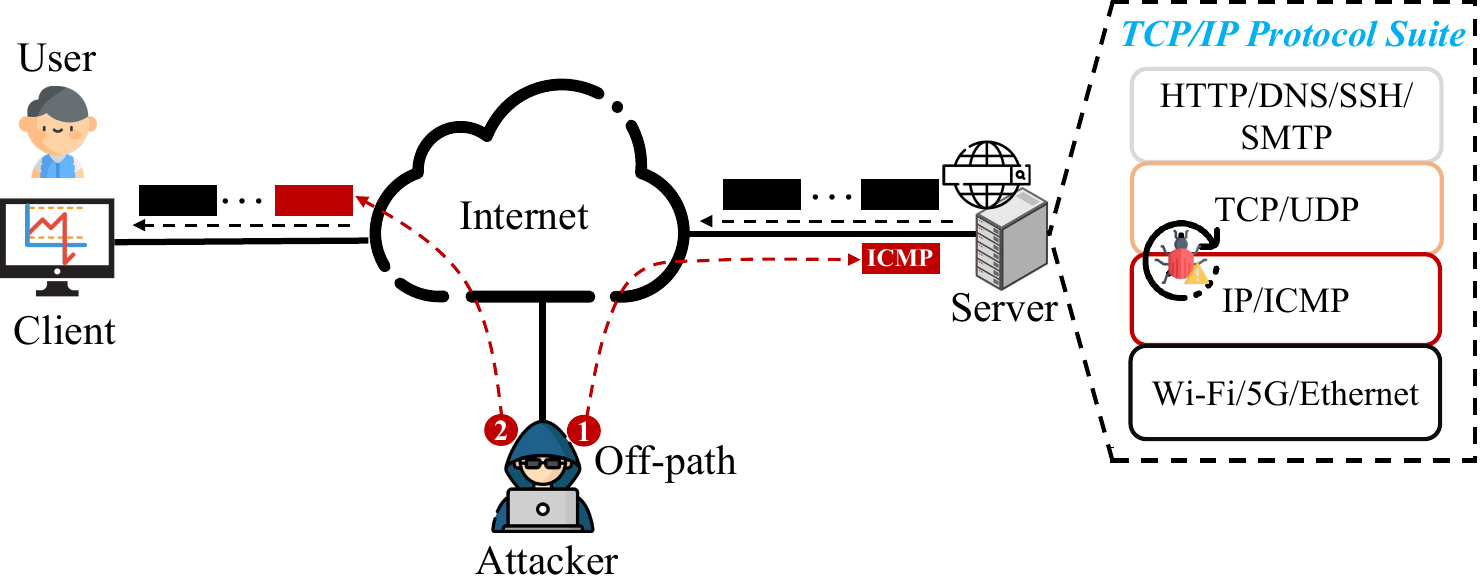}
		\vspace{-2mm}
		\caption{Threat model of off-path attacks on the TCP/IP protocol suite via forged ICMP error messages.}
		\label{pic:overview}
	\end{center}
	\vspace{-4mm}
\end{figure}

\xw{
Figure~\ref{pic:overview} shows the threat model of our off-path attacks on the TCP/IP protocol suite via forged ICMP error messages.
The off-path attacker is positioned outside the direct communication path between the server and the client. Consequently, the attacker cannot intercept or directly modify packets in transit between the server and the client.
Instead, the attacker can forge and send packets with arbitrary source IP addresses\footnote{Prior studies show that about a quarter of ASes on the Internet do not filter packets with spoofed source addresses leaving their networks, and it is trivial to rent such a machine from a bulletproof hosting node~\cite{luckie2019network,ccsman,lichtblau2017detection}}. 
Specifically, by leveraging forged ICMP error messages, the attacker exploits weaknesses and forces exceptional behaviors during cross-layer interactions among multiple protocols within the server's TCP/IP protocol suite.
Once these vulnerabilities are triggered, network traffic from the server to the client will be affected. Furthermore, the off-path attacker can impersonate the server and inject crafted packets to the client to manipulate the target network traffic.}

\xw{
The following four sections delve into four of our works that identified vulnerabilities (information leakage~\cite{ccsfeng}, desynchronization~\cite{feng2022ndss}, semantic gap~\cite{feng2022off-redirect}, and identity spoofing~\cite{feng2022man}) caused by forged ICMP error messages, enabling off-path attackers to launch impactful attacks.
}

\section{Information Leakage}
\label{sec:IPID}

TCP plays a fundamental role within the TCP/IP protocol suite and carries significant importance on the Internet. It ensures that data packets reach their intended destinations accurately and in the correct sequence.
Regarding security, the 32-bit randomization of sequence and acknowledgment numbers within the TCP protocol is a pivotal measure. \xw{This randomization strengthens the protocol's resilience against out-of-band malicious TCP packet injections.}
However, despite the extensive randomized sequence and acknowledgment number space, which \xw{significantly increases the time needed for brute force attacks}, TCP protocol operations involve interactions with other protocols in the TCP/IP protocol suite.
During these interactions, certain fields of other layer protocols (such as the IPID field of the IP protocol) can be exploited to infer the TCP protocol's sequence and acknowledgment numbers.
Particularly, we discovered that the IPID field, even with the most advanced IPID assignment policy currently available in Linux systems, can be manipulated by a forged ICMP error message issued by off-path attackers.
This manipulation allows the attacker to indirectly infer confidential information (i.e., the sequence and acknowledgment numbers of TCP) by observing the IPID field, ultimately leading to information leakage during protocol cross-layer interactions. This can enable the off-path attacker to inject malicious TCP packets into the target connection, thereby jeopardizing the integrity of the associated TCP stream~\cite{ccsfeng}.

\subsection{IPID Assignment}

The \texttt{Identification} field of IP protocol (IPID) is used to enable de-fragmentation. After abandoning previous vulnerable IPID assignment methods (e.g., global IPID assignment and per-destination IPID assignment), modern OSes typically employ advanced methods to assign IPIDs for IP packets.
For instance, Linux systems use a per-socket-based IPID assignment policy for TCP packets and utilize 2048 globally shared hash counters for non-TCP packets~\cite{alexander2019detecting}.
While this IPID assignment method aims to safeguard TCP protocols against information leakage stemming from IPID values, we demonstrate its vulnerability that can be exploited by off-path attackers to deduce the upper-layer sequence and acknowledgment numbers of a victim TCP connection.

\begin{figure*}[h]
\vspace{-6mm}
	\begin{center}
		\subfigure[The specified sequence number is wrong.]{ 
			\label{pic:seq_no}  
			\includegraphics[width=0.38\textwidth]{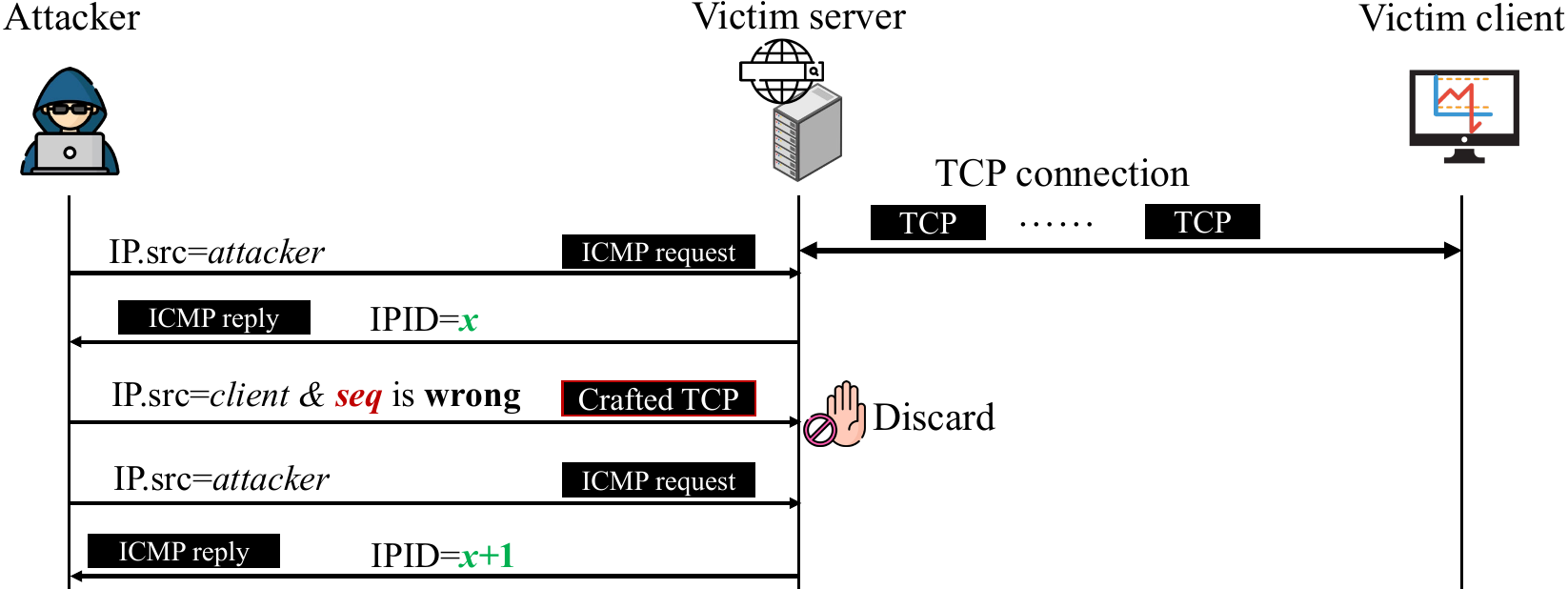} 
		} 
		\subfigure[The specified sequence number is correct.]{ 
			\label{pic:seq_yes}
			\includegraphics[width=0.38\textwidth]{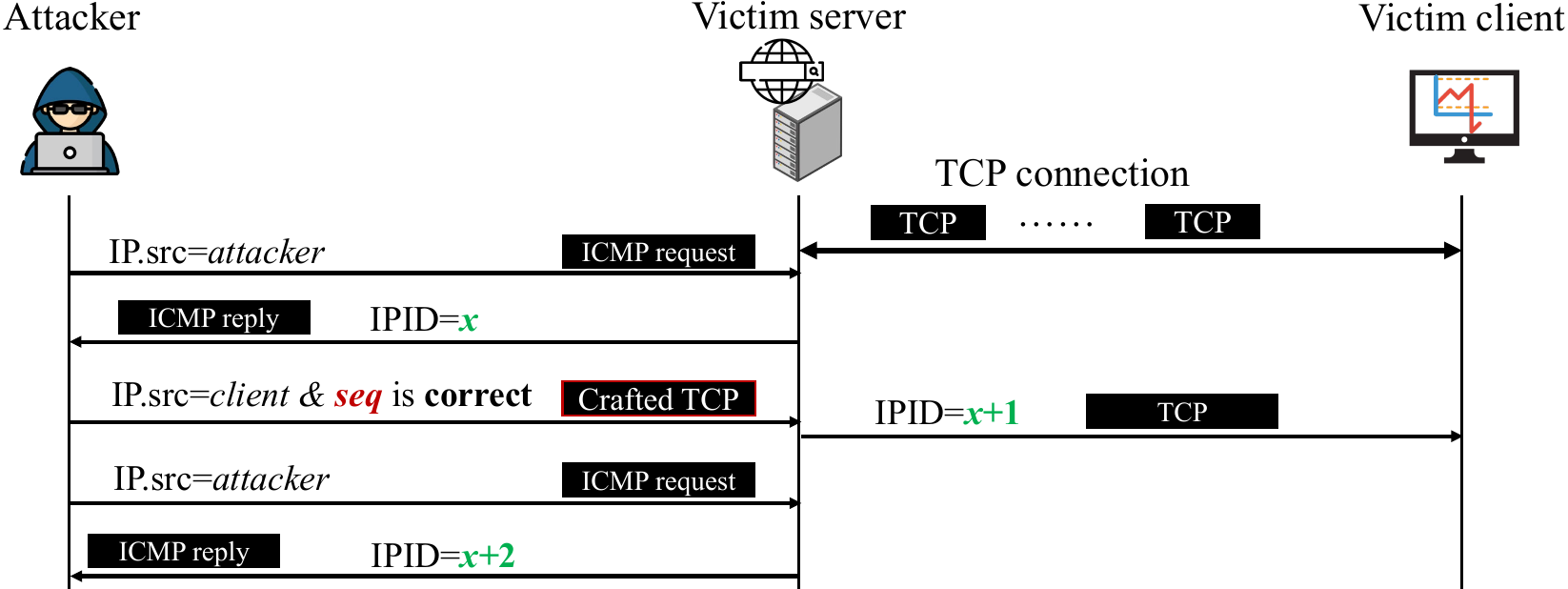}
		}
		\vspace{-4mm}
		\caption{The attacker determines the accuracy of the specified sequence number by observing the shared IPID counter~\cite{ccsfeng}.} 
		\label{pic:seq_infer} 
	\end{center}
	\vspace{-5mm}
\end{figure*}

\subsection{Inference of the Randomized Numbers}

In this situation, the off-path attacker pretends to be a router and issues a \xw{crafted ICMP error message (an ICMP Destination Unreachable message with the code `Packet too big') embedded with a 28-octet payload of a fake ICMP echo reply packet} to a Linux server. This crafted ICMP error message evades the server's legitimacy check and deceives it into downgrading its IPID assignment policy for TCP packets. The policy transitions from a per-socket-based policy to the utilization of 2048 globally shared hash counters.
Given the limited size of the hash counter pool (i.e., 2048), the attacker can change its IP address to successfully provoke a hash collision with a victim TCP client of the server\footnote{Since kernel version 5.12.4, Linux has used a dynamic hash counter pool proportional to physical RAM size to mitigate IPID-based firewall attacks~\cite{klein2022subverting}.}. This occurs because Linux servers select one of the 2048 hash IPID counters based on the destination IP address of outgoing packets.
Consequently, the server can be tricked into selecting the same IPID counter for both the attacker's IP address and the victim client's IP address. This situation allows the attacker to deduce the specific IPID counter being utilized by the Linux server for the victim TCP connection, thus creating a side channel through which information of the victim TCP connection can be leaked.

Once the shared IPID counter is identified, the attacker proceeds to send crafted TCP packets to the victim server. The shared IPID counter will exhibit varying behaviors under different circumstances, enabling the attacker to discern whether the specified values in the forged TCP packets are correct or not.
As shown in Figure~\ref{pic:seq_infer}, the attacker initiates the process by sending an ICMP echo request packet to the server and monitors the current value of the shared IPID counter when the server responds with a reply packet.
Subsequently, the attacker impersonates the identity of the victim client by IP spoofing and crafts a TCP packet destined for the server. This crafted packet includes the specified sequence number (i.e., $seq$). In this scenario, the server's behavior varies based on the sequence number specified within the crafted packet.
As illustrated in Figure~\ref{pic:seq_no}, when the specified sequence number is incorrect (i.e., not within the server's receive window), the server simply discards the packet. When the attacker later observes the current value of the shared IPID counter once more, it will notice that the IPID counter's values remain consecutive.

On the contrary, if the specified sequence number is correct (as shown in Figure~\ref{pic:seq_yes}), the server will generate a reply packet destined for the client, even though the client will ultimately discard this reply. This reply packet consumes a value from the shared IPID counter. When the attacker subsequently observes the current value of the shared IPID counter once more, it will notice that the IPID counter's values are no longer consecutive. By making this comparison, the attacker can accurately deduce sensitive information, such as sequence and acknowledgment numbers of the target TCP connection.
\xw{For off-path TCP injection attacks (as described in~\cite{cao2016off} and \cite{qian2012off}), once off-path attackers identify the randomized sequence and acknowledgment numbers of the target TCP connection, they can craft an out-of-band TCP packet specified with these identified numbers in the TCP header. When injected into the target connection, this packet will pass verification and be accepted by the receiver, potentially terminating or poisoning the connection.}

\subsection{Experimental Results}

\begin{figure}[h]
    \vspace{-2mm}
	\begin{center}
		\includegraphics[width=0.29\textwidth]{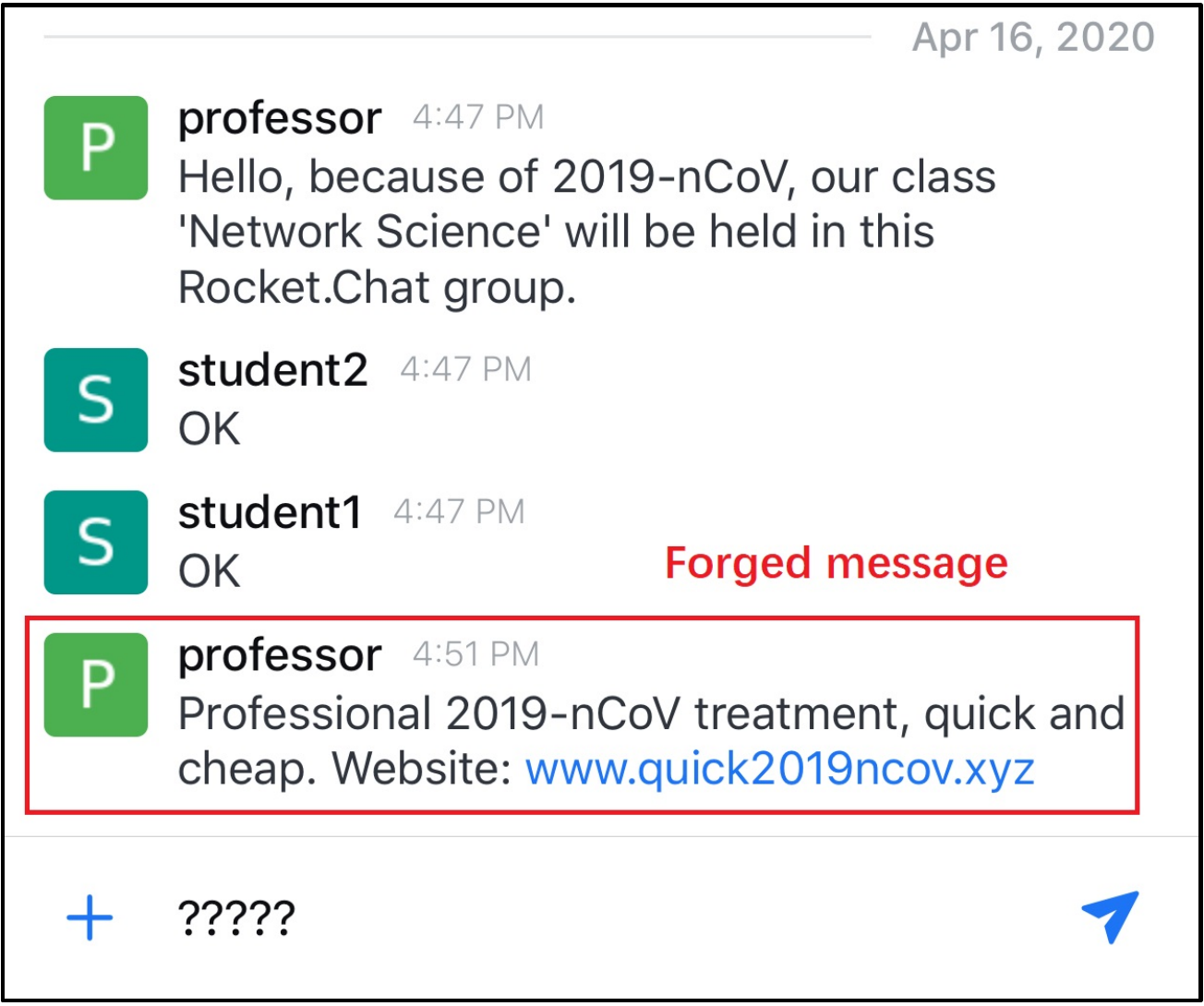}
		\vspace{-3mm}
		\caption{Snapshot of web application poisoning~\cite{ccsfeng}.}
		\label{web-poisoning}
	\end{center}
	\vspace{-4mm}
\end{figure}

Through real-world evaluations, we demonstrate that the information leakage caused by the cross-layer interactions can lead to severe real-world consequences. It can enable attackers to infer and disrupt a large number of TCP connections.
We observe that more than 20\% of the Alexa top 100k websites are vulnerable.
\xw{We establish a TCP connection from our client to each of the websites in the Alexa top 100k list. Then, an attack machine on the Internet issues a forged ICMP `Packet too big' to the website to manipulate its IPID assignment for our client. Our experimental results show that 20\% of the websites can be tricked into downgrading the IPID assignment from the per-socket-based policy to the hash-based policy for their TCP packets after receiving forged ICMP error messages.}
We implement a prototype and perform case studies on a wide range of applications, e.g., HTTP, SSH and BGP, to validate the effectiveness of the identified off-path TCP hijacking attack due to the cross-layer information leakage.
\xw{We demonstrate that an off-path attacker can infer the sequence number of a target TCP connection on port 22 within 155 seconds, thus crafting a out-of-band TCP \texttt{RST} packet to tear down the victim SSH session to cause a DoS attack. Besides, the attacker can infer the sequence and acknowledgment numbers of a target TCP connection within 215 seconds, thus crafting a TCP data packet to poison web applications or BGP routing tables~\cite{ccsfeng}.
}
%
Figure~\ref{web-poisoning} illustrates a snapshot of our attack against web applications, in which an attacker identifies a TCP connection and proceeds to inject a fake message into that connection.

\section{Desynchronization}
\label{sec:fragmentation}

Desynchronization within the TCP/IP protocol suite caused by crafted ICMP errors refers to a situation where multiple protocols simultaneously work with the same variable or data unit. Factors such as network delays or conditional competition introduced by a crafted ICMP error message can cause these protocols to lose synchronization, leading to ambiguity regarding the value of that variable or data unit. This disruption can degrade the network's original functionality or semantics, creating opportunities for attackers to exploit and compromise network systems.

Specifically, consider the path MTU value, which is a global variable maintained within the host's IP layer. This value defines the maximum IP packet size for the path from the host to a specific destination IP address.
Operations on the path MTU value extend beyond the IP protocol and involve various other protocols like TCP and UDP. Ideally, using the path MTU value to determine TCP segment size should eliminate the need for IP fragmentation.
However, we demonstrate that in practice, simultaneous updates on this global variable by various protocols, tricked by a crafted ICMP error message, can lead to desynchronization issues~\cite{feng2022ndss}. This can result in discrepancies between the path MTU value at the IP layer and the MTU value read by the TCP layer, potentially causing the TCP layer to transmit oversized segments, leading to abnormal IP fragmentation. Consequently, off-path attackers can inject manipulated IP fragments into the target TCP connection, causing mis-reassembling of IP fragments and disrupting the target TCP traffic without needing to infer random sequence numbers.

\begin{figure}[h]
    \vspace{-4mm}
	\begin{center}
		\includegraphics[width=0.44\textwidth]{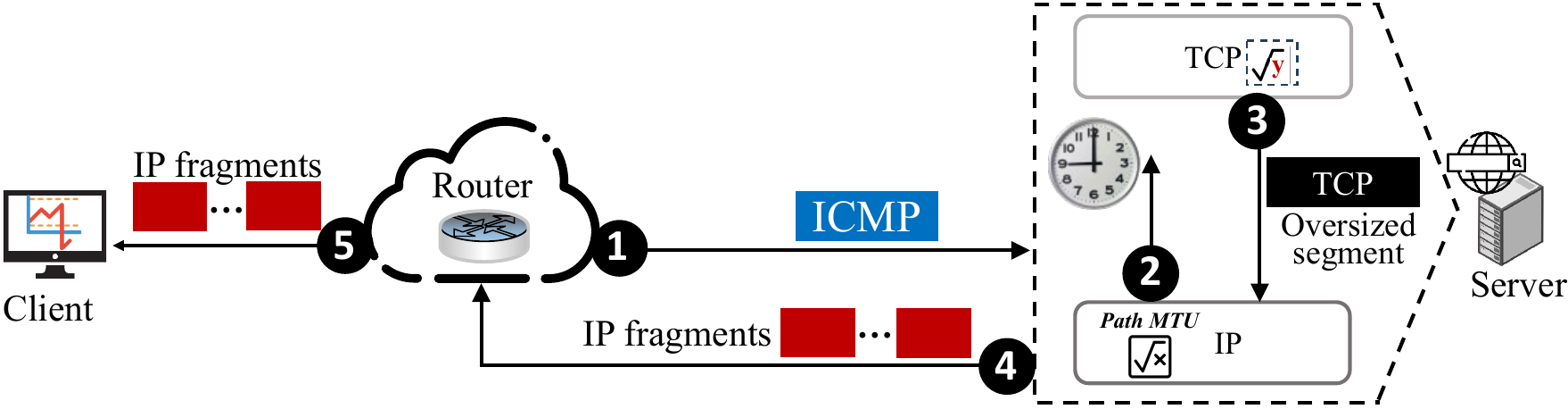}
		\vspace{-2mm}
		\caption{IP fragmentation on TCP segments due to the desynchronization on path MTU value between IP and TCP.}
		\label{pic:fragmentation}
	\end{center}
	\vspace{-6mm}
\end{figure}

\subsection{Forcing IP Fragmentation on TCP}

It is a widespread belief that TCP is immune to IP fragmentation because TCP enables the Path MTU Discovery (PMTUD) mechanism by default. This mechanism detects the maximum allowed packet size along the path and enables TCP to adjust the maximum segment size (MSS) accordingly, thus avoiding IP fragmentation on TCP segments.
In practice, the detected path MTU value is a global variable maintained at the IP layer. Consequently, when multiple protocols, such as IP, TCP, UDP, and others, simultaneously interact with it, unexpected synchronization issues may occur, resulting in unintended IP fragmentation on TCP segments.

As shown in Figure~\ref{pic:fragmentation}, a router on the Internet may generate an ICMP error message (\xw{an ICMP Destination Unreachable message with the code `Packet too big'}) directed at the server. This ICMP error message can be triggered by various protocol sessions from the server, such as UDP or ICMP echo. Upon reaching the server, this message updates the global variable of path MTU in the IP layer based on its contents. However, as this message lacks specific TCP connection information, the update to the path MTU value is not immediately synchronized with the TCP layer. \xw{Instead, the IP layer defers feedback until it passively detects the TCP connection by receiving oversized TCP segments, which it then fragments and sends out. Once the IP layer acknowledges the TCP connection, it updates the TCP layer with the new PMTU, allowing TCP to adjust the MSS of subsequent segments to avoid IP fragmentation.}

This desynchronization issue concerning the path MTU value between TCP and IP undermines the primary purpose of the path MTU discovery mechanism and causes unintended IP fragmentation on TCP segments.
Particularly, we find that off-path attackers on the Internet can impersonate a router and forge such an ICMP error message to trick the server into fragmenting its TCP segments. This manipulation exploits the inherent challenge in verifying the source and transmission path of ICMP error messages within the current Internet infrastructure. For example, we can forge the ICMP error message to include an embedded ICMP echo reply packet, effectively tricking the server into fragmenting TCP segments and introducing a new attack vector.

\subsection{Poison TCP Traffic via IP Fragmentation}

\begin{figure}[h]
    \vspace{-3mm}
	\begin{center}
		\includegraphics[width=0.44\textwidth]{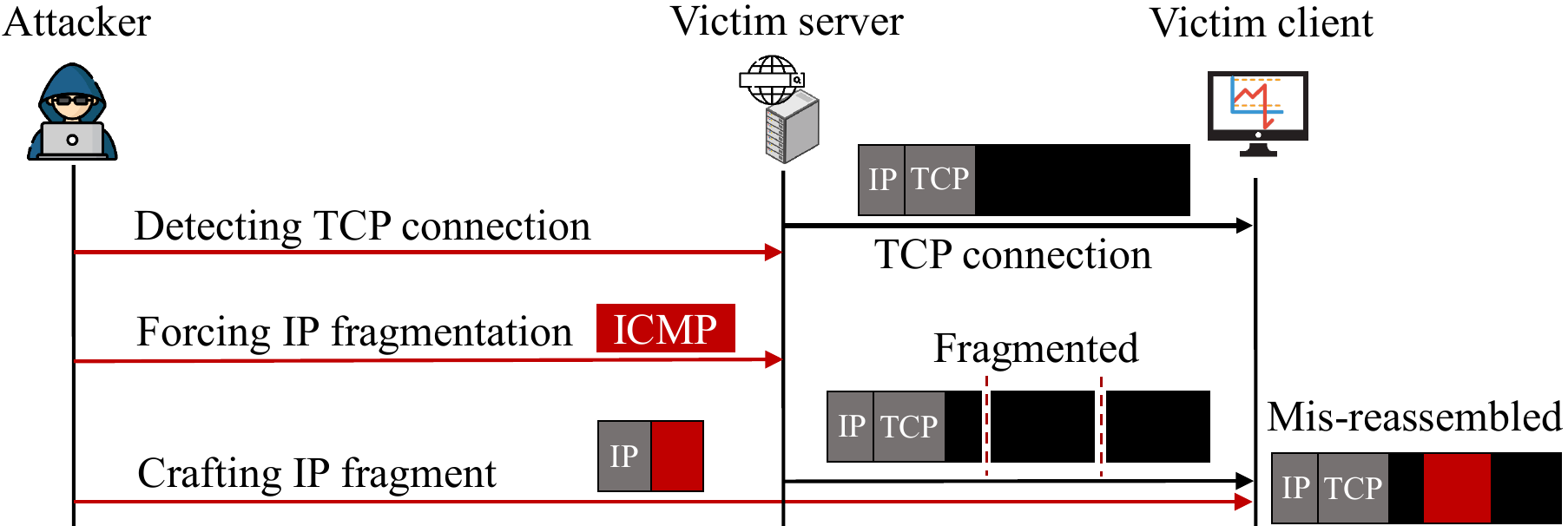}
		\vspace{-2mm}
		\caption{Poisoning TCP traffic via IP fragmentation~\cite{feng2022ndss}.}
		\label{pic:frag_poison}
	\end{center}
	\vspace{-6mm}
\end{figure}

Once TCP packets experience IP fragmentation due to the desynchronization issue, an off-path attacker may exploit this vulnerability to launch IP fragmentation injection attacks against TCP traffic.
As shown in Figure~\ref{pic:frag_poison}, at first the off-path attacker may employ various techniques, such as social engineering or network-side channels, to detect the existence of a TCP connection between a victim server and client. Subsequently, the attacker forges an ICMP error message and sends it to the server, triggering the desynchronization vulnerability on path MTU in the server's TCP/IP protocol suite. This manipulation causes IP fragmentation on the TCP packets sent from the server to the client.
Following this, the attacker impersonates as the server via IP spoofing and sends crafted IP fragments to the victim client.
As a result, legitimate fragments from the server will be incorrectly reassembled with the malicious ones introduced by the attacker. Ultimately, this leads to the replacement of the original data within the TCP packets, initiating a poisoning attack on the targeted TCP stream.
\xw{According to RFC 791, the minimal IP fragments on the Internet is 68 octets, thus the random sequence and acknowledgment numbers are always carried in the first benign fragment from the server. Consequently, IP fragmentation-based poisoning attack against TCP can be performed without the need to infer the random sequence and acknowledgment numbers.}\footnote{\xw{It is worth noting that the handling of overlapped IP fragments is an implementation decision. Popular operating systems (e.g., Linux, OpenBSD, Windows) handle overlapped IP fragments on a first-come, first-served basis~\cite{feng2022ndss}, which allows attackers to send crafted IP fragments to the victim client in advance, facilitating the construction of our attack.}
}

\subsection{Experimental Results}

%

We demonstrate that off-path attackers can manipulate HTTP traffic via our attack. \xw{A malicious JavaScript installed at the victim client due to spam aids the attacker in synchronizing timing and aligning data to poison the local web cache\footnote{The malicious JavaScript is sandboxed by the client's browser, having limited privileges and cannot access any information within the TCP/IP protocol suite~\cite{feng2022ndss}.}. The connection to the target HTTP server is established by the puppet, and the connection and segments from the HTTP server are known to the attacker. Consequently, leveraging our method, the attacker can craft subtle IP fragments to force the incorrect reassembly of both legitimate and malicious fragments, thereby poisoning the client's web cache, leading to regular users encountering poisoned local cached data when accessing the HTTP server later.}

%

\begin{figure}[h]
    \vspace{-3mm}
	\begin{center}
		\includegraphics[width=0.35\textwidth]{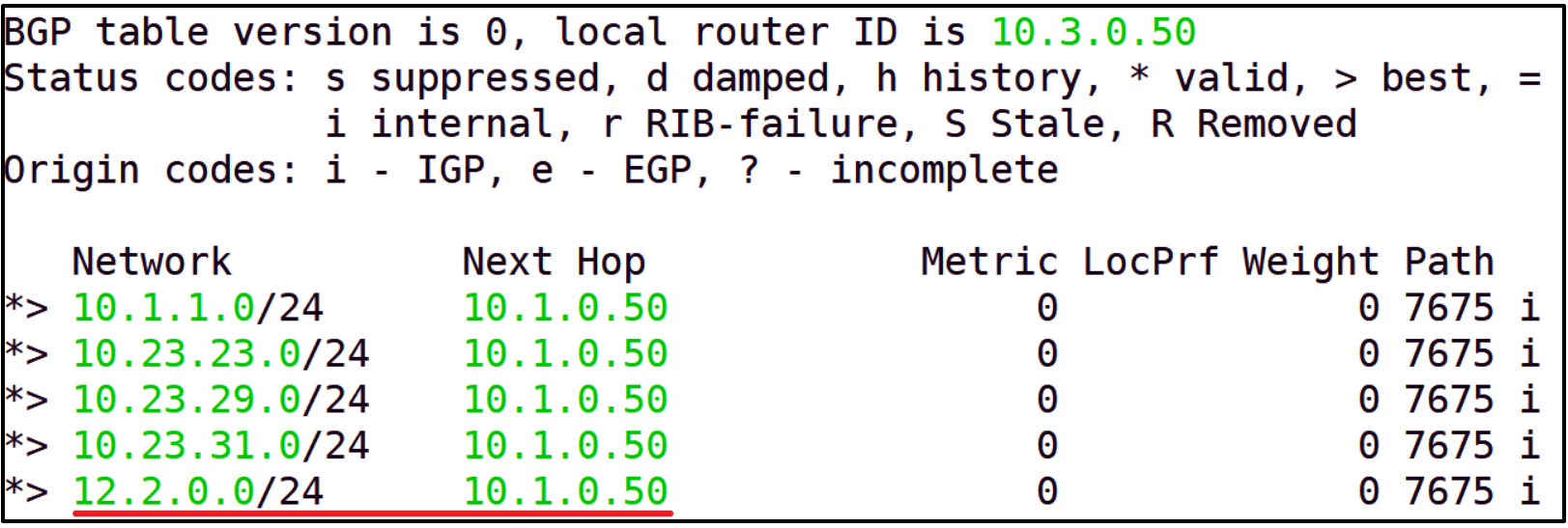}
		\vspace{-3mm}
		\caption{Fake routing due to IP fragments injection~\cite{feng2022ndss}.}
		\label{pic:fake_routing}
	\end{center}
	\vspace{-5mm}
\end{figure}

Furthermore, we demonstrate that an off-path attacker can manipulate BGP routing tables via our attack. The attacker first probes periodically advertised BGP messages in advance~\cite{flavel2010bgp}. Then, it manipulates BGP routers into fragmenting TCP segments by sending forged ICMP error messages. Finally, the attacker injects forged fragments into the BGP messages to poison the routing tables.
As illustrated in Figure~\ref{pic:fake_routing}, we present the altered routing information received by a victim BGP router within our test-bed environment, which differs from the original routing information advertised by its peer BGP router. In this scenario, the attacker has replaced the legitimate routing information of 10.2.2.0/24 with a counterfeit entry of 12.2.0.0/24 by injecting meticulously crafted IP fragments into the victim BGP router.
Our experimental findings indicate that these attacks can inflict significant threat to the Internet infrastructure. 
%
\xw{It is worth noting that session encryption mechanism (e.g., TLS) will mitigate the identified IP fragmentation attacks against TCP, since the mis-reassembled TCP segment cannot pass the up-layer verification and will be discarded. However, the discarding of the mis-reassembled TCP segment will incur performance loss, since benign fragments are also discarded together.}

\section{Semantic Gap}
\label{sec:redirect}

The concept of semantic gap signifies that protocols may inherently fall short in comprehensively addressing the wide spectrum of data types and exceptional scenarios during the processing of received packets carrying cross-layer data. This deficiency can give rise to undefined or inadequately considered situations. To maintain network functionality, protocols may resort to employing default and imprecise processing methods when responding to these packets, thereby introducing the possibility of semantic mismatches that attackers could potentially exploit to compromise the system's security. 
Specifically, we uncover that due to such a semantic gap vulnerability in the legitimacy checks against ICMP error messages, an off-path attacker on the Internet can craft an ICMP redirect message to evade the receiver's (e.g., a public server) checks. This tricks the receiver into modifying its routing table incorrectly and forwarding its IP traffic to blackholes, i.e., conducting a stealthy DoS attack against public servers on the Internet~\cite{feng2022off-redirect}.

\subsection{DoS via Semantic Gap of ICMP Error Checks}

\begin{figure}[h]
    \vspace{-3mm}
	\begin{center}
		\includegraphics[width=0.42\textwidth]{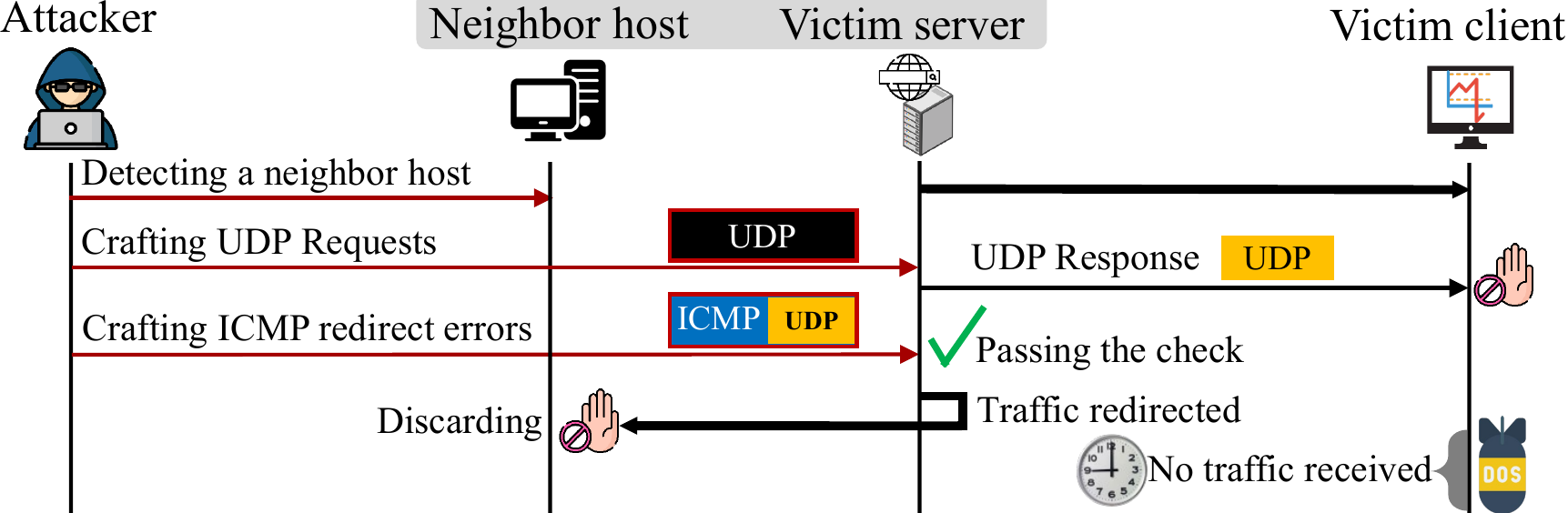}
		\vspace{-2mm}
		\caption{DoS via semantic gaps of ICMP error checks~\cite{feng2022off-redirect}.}
		\label{pic:DoS}
	\end{center}
	\vspace{-6mm}
\end{figure}

Figure~\ref{pic:DoS} illustrates our design for constructing a DoS attack against the victim server, redirecting its traffic intended for the victim client into a blackhole hosted by a neighboring host of the server that lacks the capability to forward network traffic.
Initially, the server can successfully forward its traffic to the client.
An off-path attacker on the Internet identifies a neighboring host near the target server through actions like ICMP echo requests (e.g., the ping tool). This host will later serve as a routing blackhole. 
The attacker then impersonates the victim client by IP spoofing and sends a crafted UDP request to the server. Deceived by the request, the server responds with a UDP reply to the victim client, which the victim client eventually discards.

The attacker then embeds the predictable UDP reply packet into a crafted ICMP redirect message and sends it to the victim server.
\xw{According to the ICMP specifications~\cite{rfc792, rfc1122, rfc1812}, the server will check the first 28 octets of the embedded UDP reply packet to validate the legitimacy of the received ICMP redirect message, i.e., verifying the existence of the corresponding UDP socket, even though it cannot check further information due to UDP's stateless nature.
Since the attacker previously tricked the server into establishing the UDP socket for the victim client, this crafted ICMP redirect message will pass the server's legitimacy check.}
Consequently, the server will mistakenly accept and respond to the message, redirecting its traffic for the victim client to the neighboring host as specified by the forged ICMP redirect message. However, the neighboring host lacks routing and forwarding capabilities and will discard the server's traffic. This results in a cross-layer DoS attack on all network sessions above the IP layer of the server.

\subsection{Experimental Results}

\begin{figure}[h]
    \vspace{-2mm}
	\begin{center}
		\includegraphics[width=0.38\textwidth]{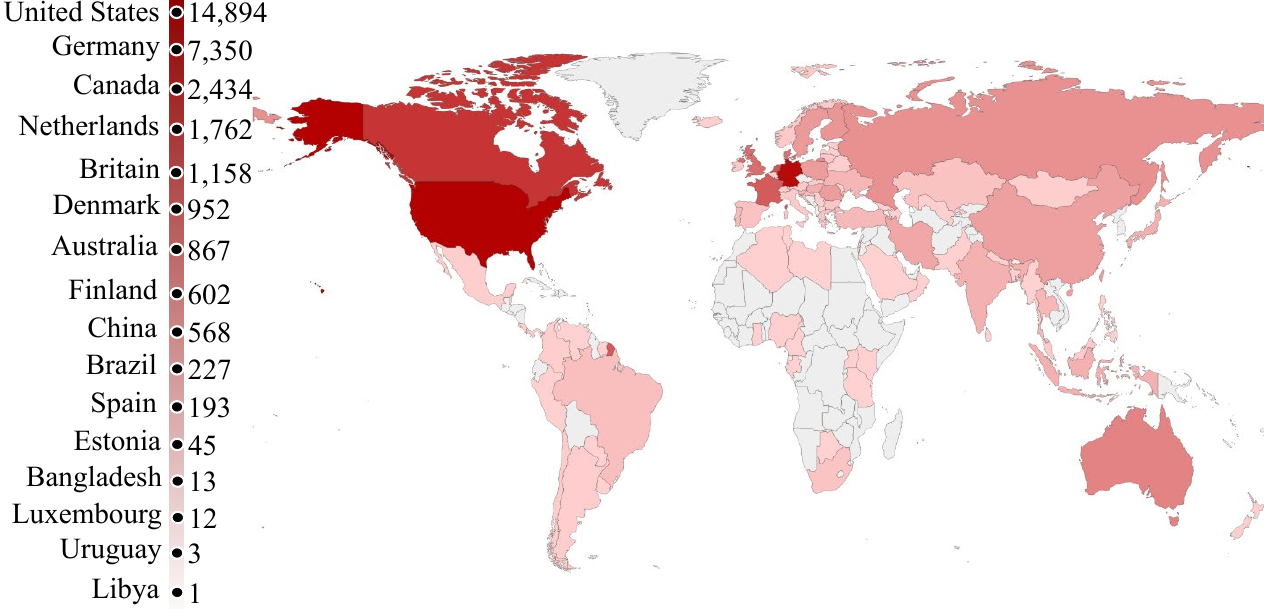}
		\vspace{-4mm}
		\caption{Distribution of vulnerable websites~\cite{feng2022off-redirect}.}
		\label{location}
	\end{center}
	\vspace{-6mm}
\end{figure}

We conduct large-scale measurements on the Internet, revealing that this DoS attack, due to the semantic gap vulnerability in ICMP error message's legitimate check mechanism, can be exploited to pose a significant threat to the Internet.
\xw{In our ethical measurement studies, we first initiate a session between our controlled client and the target server (e.g., an HTTP session from our web client to a public HTTP server). Then, using the identified ICMP redirect DoS attack, we redirect the server-to-our-client traffic to a black hole.
This causes subsequent requests from our controlled client to the server to fail and demonstrates that the server is vulnerable to our attack, while not affecting regular users of the server}.
Our experimental results show that the identified DoS attack can target not only individual users, preventing them from visiting a web server, but also server-to-server communication, such as shutting down a DNS resolver from contacting a particular authoritative name server (under our control in the experiments due to ethical considerations) to resolve domain names. It is even possible to disrupt the entire operation of a service such as Tor by breaking down the communication between a Tor relay node and a next hop.
Our one-month empirical study on the Internet reveals that 43,081 popular websites, 54,470 open DNS resolvers, and 186 Tor relay nodes, spanning 5,184 ASes across 185 countries, are vulnerable to the semantic gap vulnerability and susceptible to the identified remote DoS attack.
Figure~\ref{location} shows the geographical distribution of the vulnerable websites we detected.

\section{Identity Deception}
\label{sec:Wi-Fi}

The problem of identity deception stems from the lack of security auditing for data sources during cross-layer interactions among multiple protocols within the TCP/IP protocol suite, particularly the source of ICMP errors. This allows attackers to craft specific control protocol data, disrupting the normal operation of the network.
We show that in specific network scenarios, such as Wi-Fi networks, identity deception can be particularly severe, presenting one of the most common challenges~\cite{feng2022man}.
In public Wi-Fi networks found in airports, coffee shops, campuses, hotels, etc., an attacker (a malicious client) may connect to the network and employ source IP address spoofing techniques to impersonate as the AP gateway.
\xw{
The attacker can then send forged ICMP routing updates control messages (i.e., ICMP redirect messages designed for AP gateways only in Wi-Fi networks) to other clients. These forged ICMP routing updates control messages will pass through the AP gateway, if the AP gateway fails to block these forged messages which finally arrived at other clients, these clients will be tricked into following the message's instruction and setting the attacker as their new AP router, granting the attacker the ability to intercept wireless traffic within the Wi-Fi network.}
What is more concerning is that this vulnerability enables the attacker to evade Wi-Fi protocol security measures like WPA3, granting access to plaintext traffic within the Wi-Fi network.

\subsection{Wi-Fi Traffic Hijacking}

\begin{figure*}[h]
\vspace{-9mm}
	\begin{center}
	\subfigure[The sniffed encrypted frames.]{ 
		\includegraphics[width=0.25\textwidth]{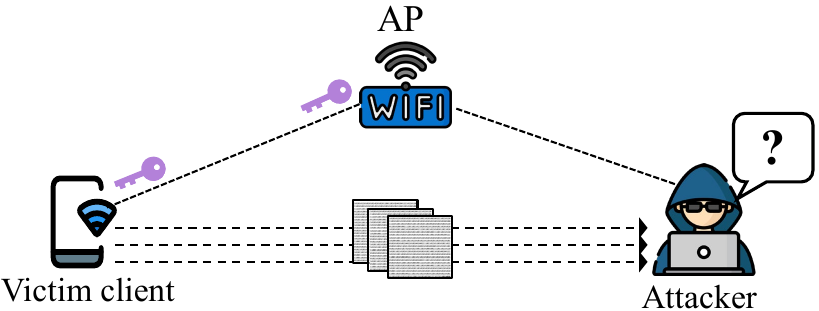} 
	} 
	\subfigure[Forged ICMP error to the victim.]{ 
		\label{pic:forging_redirec_2}
		\includegraphics[width=0.25\textwidth]{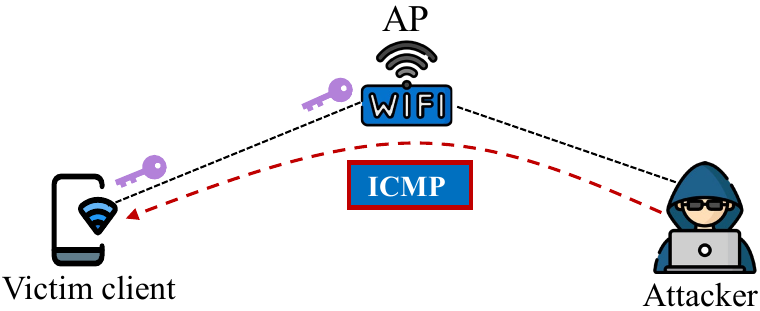}
	}
	\subfigure[Plaintext intercepted by the attacker.]{
		\includegraphics[width=0.25\textwidth]{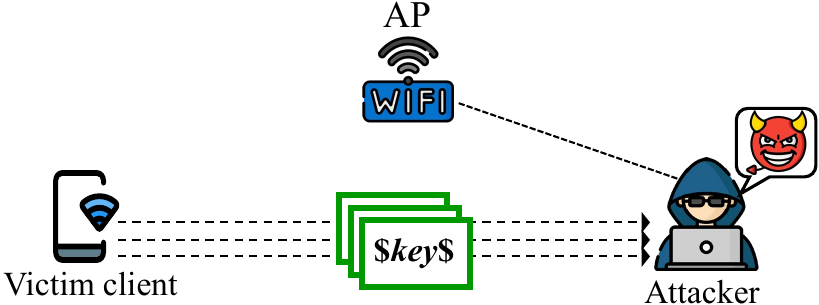}
	}
	\vspace{-5mm}
	\caption{Wi-Fi traffic interception via identity deception.}
	\label{pic:wi-fi} 
	\end{center}
\vspace{-5mm}
\end{figure*}

Figure~\ref{pic:wi-fi} shows the overview of how to intercept plaintext traffic in Wi-Fi networks by leveraging the vulnerability of identity deception.
In Wi-Fi networks, due to the shared nature of wireless channels, a malicious client (i.e., an attacker) may eavesdrop wireless frames belonging to other clients. However, these frames are usually encrypted by security mechanisms at the link layer, such as WPA2 or WPA3. As a result, it is difficult for the attacker to directly access plaintext information.
We discover a security vulnerability within the Network Processing Unit (NPU) employed in AP routers. Driven by the quest for high-speed packet forwarding, these NPU chips within AP routers directly forward received ICMP messages (including forged ICMP errors from the attack) at the hardware level, thus failing Access Control List (ACL) rules defined at the higher layers to verify and block forged messages.

As shown in Figure~\ref{pic:forging_redirec_2}, this vulnerability allows the attacker to impersonate the AP router and craft an ICMP redirect message to manipulate the IP routing of the victim client. Even though such a message is meant to exclusively originate from the AP router itself and exhibits obvious illegitimate characteristics (e.g., its source being the AP router's IP address), due to the NPU's direct forwarding of the message, this message passes through the AP router and remains unblocked.
Ultimately, upon reaching the victim client, this message deceives the client into perceiving it as originating from the AP router. Subsequently, the victim client updates and alters its IP routing based on the message's functional requirements, designating the attacker as the next-hop gateway in its IP layer routing. Consequently, all ensuing traffic from the victim client is rerouted through the attacker. 
%

%
Note that this attack can bypass the link-layer encryption protection mechanisms employed in Wi-Fi networks (e.g., WPA2 and WPA3).
\xw{WPA2 and WPA3 provide per-hop encryption at the link layer using a session key shared between the AP and each attached client. However, due to the crafted ICMP redirect message, the victim client sets the attacker as the next hop in the IP layer. Therefore, when the AP receives the encrypted link-layer frames from the victim client, it needs to perform multi-hop relaying at the link layer to complete forwarding the frames to the next hop (i.e., the attacker).
Consequently, the AP first decrypts the encrypted frames using the shared secret key with the victim client. Next, according to the \texttt{Destination Address} (which has been poisoned as the attacker) in the frame header, the AP encrypts the frames using the secret key shared with the attacker and sends them to the attacker. Finally, after decrypting the frames, the attacker can intercept the victim client’s plaintext traffic, and the link-layer per-hop encryption in Wi-Fi networks is successfully evaded~\cite{feng2022man}.
}

\begin{figure}[h]
	\vspace{-3mm}
	\begin{center}
		\includegraphics[width=0.3\textwidth]{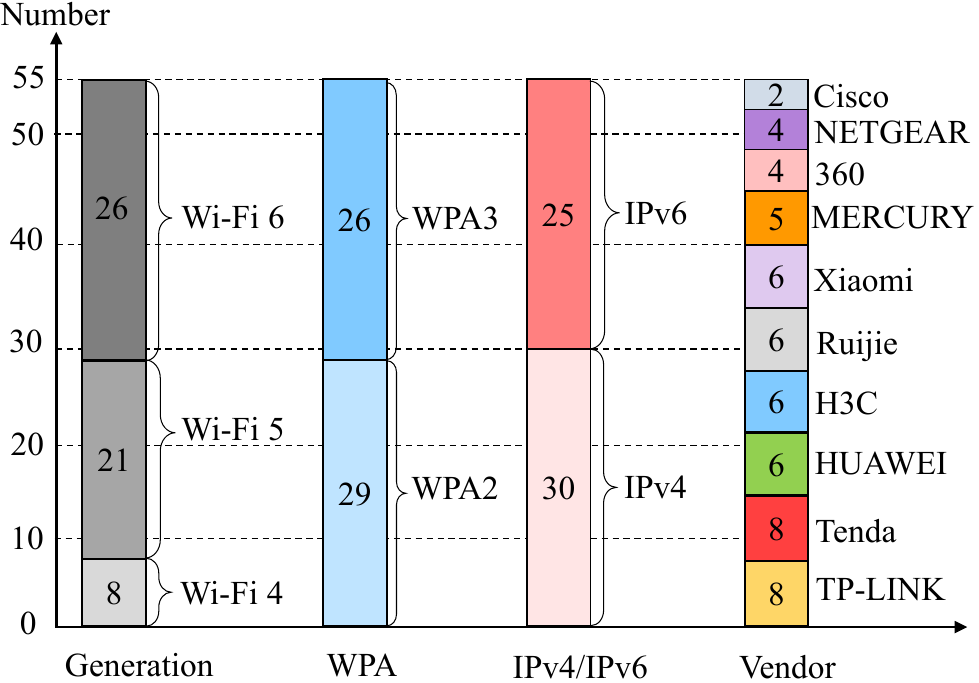}
		\vspace{-3mm}
		\caption{Distribution of 55 vulnerable AP routers~\cite{feng2022man}.}
		\label{pic:AP-routers}
	\end{center}
	\vspace{-5mm}
\end{figure}

%

\subsection{Experimental Results}

We conduct real-world evaluations to assess the impact of our attack. Initially, we investigate whether popular AP routers could effectively block forged ICMP redirect messages sent from an attacker to a victim client. Our assessment covers 55 popular wireless routers spanning 10 vendors (as shown in Figure~\ref{pic:AP-routers}). Our findings reveal that none of these routers block forged ICMP redirects passing through.
The root cause of this identity deception vulnerability, stemming from the flawed design of NPUs, has been officially recognized by Qualcomm (CVE-2022-2566) and HiSilicon (HWSA21-085272813). Additionally, HUAWEI, H3C, Ruijie, MERCURY, NETGEAR, and Tenda have also confirmed the presence of this vulnerability in their AP routers due to the NPU.
Furthermore, we evaluate 122 real-world Wi-Fi networks, encompassing various scenarios such as those in coffee shops, hotels, libraries, cinemas, campuses. Our findings reveal that 109 of these networks, equating to 89\%, are vulnerable to our traffic hijacking attack.

\section{Countermeasures}
\label{sec:countermeasure}

\xw{We responsibly disclosed the identified vulnerabilities to the affected organizations. We reported the IPID assignment policy vulnerability, triggered by a forged ICMP `Packet Too Big' message, to the Linux community. They acknowledged it (CVE-2020-36516) and improved the IPID design starting from kernel version 5.16. The desynchronization and semantic gap vulnerabilities, which are exploitable for IP fragmentation and remote DoS attacks, were reported to Linux and FreeBSD. Both confirmed receipt, and we are awaiting updates. Qualcomm acknowledged and fixed the Wi-Fi identity deception vulnerability caused by crafted ICMP redirects in their Snapdragon chipsets (CVE-2022-2566); other affected vendors are still working on fixes. Besides, we reported the vulnerabilities in the legitimacy check mechanism of ICMP errors to the IETF and are discussing them with our countermeasures.
}

\subsection{Enhancing ICMP Error Authentications}
%
%
%
%
\xw{
The root cause of the four off-path attacks presented in this paper is that an off-path attacker can forge ICMP error messages to bypass the receiver's legitimacy checks, leading to unintended protocol interactions and vulnerabilities. The most straightforward prevention measure is to strengthen the authentication of received ICMP error messages. However, as discussed in \S\ref{subsec:icmpbasics}, verifying the legitimacy of ICMP errors is challenging due to two inherent limitations in the current ICMP specifications.
Firstly, certain ICMP errors (e.g., ICMP Destination Unreachable messages with the code `Packet too big' exploited to trigger the information leakage and desynchronization vulnerabilities) can originate from any intermediate routers, rendering source-based blocking ineffective.
Secondly, although ICMP specifications mandate including at least the first 28 octets of the original packet, off-path attackers can evade this by embedding a crafted UDP or ICMP payload into the forged ICMP error messages, thwarting authentication due to the statelessness and lack of memory in UDP and ICMP protocols.
}

\xw{
Inspired by RFC 5961's challenge ACK mechanism~\cite{rfc5961} for defending against out-of-band TCP packet injection, we propose enhancing ICMP error authentication by introducing a new \textit{challenge-and-confirm} mechanism.
%
Particularly, when a receiver gets an ICMP error message embedded with a stateless protocol payload (like UDP/ICMP), verifying its authenticity can be difficult. To address this, the receiver can send another (UDP/ICMP) packet on the established network session to the destination, embedding a hash value in the IP options field. If the prior ICMP error message was legitimate, this new packet will trigger another ICMP error message containing the hash value. This allows the receiver to verify authenticity and respond correctly. This \textit{challenge-and-confirm} mechanism effectively defends against off-path forged ICMP error messages with minimal changes to the TCP/IP protocol suite. It only requires updates to the ICMP error message verification code on end hosts, without modifying intermediate routing devices, and it is backward compatible. We are discussing this mechanism with the IETF.}

\subsection{Securing Sessions via Cryptography}

\xw{
Another mitigation method is to use cryptography to secure network sessions as much as possible, such as with TLS~\cite{rfc8446}, QUIC~\cite{rfc9000}, and TCP-MD5/TCP-AO~\cite{rfc5925}. This way, even if an off-path attacker exploits forged ICMP error messages to trigger vulnerabilities in the TCP/IP stack, it is difficult for the attacker to cause real harm to applications.
For instance, even if an attacker manipulates the server's IPID with ICMP error messages to create a side channel and guesses the sequence number of a target TCP connection, the injected TCP packet will fail TCP-MD5/TCP-AO or TLS validation and be discarded. Similarly, if an off-path attacker intercepts a victim client's packets in a Wi-Fi network as a Man-in-the-Middle and evades link-layer encryption like WPA3, the end-to-end encryption provided by protocols such as TLS or QUIC makes it challenging for the attacker to access plaintext application data, thereby limiting the attack's impact.
}

\section{Discussion}
\label{sec:problem}

\xw{
Off-Path attacks on the TCP/IP protocol suite present a significant challenge to Internet security, as they do not constrain the attacker's network topology and require minimal resources.
Previous research has demonstrated that off-path attackers can exploit vulnerabilities in the TCP/IP protocol suite to launch various attacks, such as TCP hijacking~\cite{qian2012off,cao2016off,pan2024tcp}, routing manipulation~\cite{nakibly2012persistent,wang2020observing}, web and DNS cache poisoning~\cite{gilad2012off,gilad2013off,klein2021cross,gilad2011fragmentation}. However, off-path attacks facilitated by forged ICMP errors have received limited attention~\cite{ccsman,man2021dns}.
}

\xw{
In our study, we systematically reveal four security issues caused by forged ICMP errors: information leakage, desynchronization, semantic gaps, and identity deception.
These issues can be exploited by attackers to pose severe security threats to the Internet. Essentially, these security issues arise from the disruption of the protocol's intended communication processes and semantic integrity by forged ICMP error messages, leading to unexpected behaviors that attackers can exploit.
We term these vulnerabilities as protocol interaction semantic vulnerabilities caused by forged ICMP errors, distinguishing them from memory corruptions caused by unsafe programming practices.
Given that ICMP (including ICMPv6) is widely implemented and crucial across various TCP/IP protocol stacks, the semantic vulnerabilities caused by forged ICMP errors may extend beyond the four we have identified.
}

\section{Conclusion and Future Work}
\label{sec:conclusion}

In this paper, we investigate the security implications of cross-layer interactions within the TCP/IP protocol suite caused by forged ICMP error messages from off-path attackers. Our research unveils four semantic vulnerabilities that can be triggered by forged ICMP errors, i.e., information leakage, desynchronization, semantic gaps, and identity deception.
%
Exploitable by off-path attackers on the Internet, these vulnerabilities have the potential to manipulate network traffic, impacting numerous popular websites, DNS resolvers, Tor nodes, and public Wi-Fi networks.
We responsibly disclosed these vulnerabilities to the affected vendors and propose our countermeasures. A critical area of focus for future research is the automated identification of these semantic vulnerabilities, for example, by leveraging techniques from program analysis~\cite{fiterau2023automata,cao2019principled} and AI models~\cite{mirskyvulchecker,thapa2022transformer}.

\clearpage

\balance
\bibliographystyle{ACM-Reference-Format}
\bibliography{reference}

\end{document}